\documentclass[12pt]{article}
\pdfoutput=1
\usepackage{amsmath, amsthm, amssymb, graphicx,slashed}
\usepackage{amsfonts}
\usepackage[backend=bibtex,sorting=none,style=numeric-comp,maxnames=8]{biblatex}
\renewbibmacro{in:}{}

\usepackage[utf8]{inputenc} 
\usepackage{hyperref} 
\hypersetup{bookmarksopen=true, bookmarksnumbered=true,
 pdfstartview={FitH}, pdfborder={0 0 0}, colorlinks=true,
 linkcolor=red, linktocpage=true, citecolor=blue, urlcolor=blue,
 unicode=true, pdftitle={LovelockLifshitz},
 pdfauthor={DJ}, pdfsubject={Einstein Gauss Bonnet theory,
   Kounterterms and AMD mass}, pdfkeywords={AdS, EGB, AMD,
   Kounterterms}}

\usepackage{geometry} 
\usepackage{fancyhdr} 
\usepackage{parskip} 
\usepackage[nottoc,notlof,notlot]{tocbibind} 
\usepackage[titles]{tocloft} 

\numberwithin{equation}{section} 

\def\be{\begin{equation}}
\def\ee{\end{equation}}
\def\ba{\begin{split}}
\def\ea{\end{split}}
\def\beq{\begin{eqnarray}}
\def\eeq{\end{eqnarray}}
\def\a{\alpha}

\bibliography{LifshitzinLovelock}

\date{}

\begin{document}

\title{\Huge\textbf{Exploring Lovelock theory moduli space for Schr\"
    odinger solutions}} \author{Dileep
  P.
  Jatkar\footnote{\href{mailto:dileep@hri.res.in}{dileep@hri.res.in}}~ and
  Nilay Kundu\footnote{\href{mailto:nilaykundu@hri.res.in}{nilaykundu@hri.res.in}}
  \bigskip\\
  \small{Harish-Chandra Research Institute,} \\
  \small{Chhatnag Road, Jhunsi, Allahabad 211019, India}}

\maketitle
\thispagestyle{fancy}
\vspace{0.3cm}
\begin{abstract}
\normalsize

We look for Schr\" odinger solutions in Lovelock gravity in $D > 4$.
We span the entire parameter space and determine parametric relations
under which the Schr\" odinger solution exists.  We find that in
arbitrary dimensions pure Lovelock theories have Schr\" odinger
solutions of arbitrary radius, on a co-dimension one locus in the
Lovelock parameter space.  This co-dimension one locus contains the
subspace over which the Lovelock gravity can be written in the
Chern-Simons form.  Schr\" odinger solutions do not exist
outside this locus and on this locus they exist for arbitrary
dynamical exponent $z$.  This freedom in $z$ is due to the
degeneracy in the configuration space. We show that this degeneracy
survives certain deformation away from the Lovelock moduli space.
\end{abstract}
\newpage

\section{Introduction} \label{intro}

The Schr\" odinger solutions belong to a class of solutions
to the gravitational equations of motion which asymptotically do not
preserve the Lorentz symmetry.  They, however, do respect some
non-relativistic symmetries.  The deviation from the relativistic
symmetry is parametrized by the Schr\" odinger scaling
exponent $z$, or the dynamical exponent.

The Schr\" odinger solution was first obtained by Son\cite{Son:2008ye}
as well as Balasubramanian and Mcgreevy\cite{Balasubramanian:2008dm}.
They assumed the stress tensor consisting of the cosmological constant
term and the pressure-less dust.
The Schr\" odinger solution possesses the Galilean
boost invariance by assigning a specific transformation property to
one of the light-like directions\cite{Son:2008ye,
  Balasubramanian:2008dm}(non-relativistic metrics in higher
derivative gravity were discussed in \cite{Adams:2008zk}).

In this note we analyse this question in more detail by spanning
the entire coupling parameter space of Lovelock theories in various
dimensions.  Up to four space-time dimensions, the Lovelock
action is identical to the Einstein Hilbert action with the
cosmological constant, but from five dimensions onwards the Lovelock
action has additional Gauss-Bonnet term in the action.  This term can 
be added in four dimensions as well but being
total derivative term it does not contribute to the dynamics.  In five
and higher dimensions the Gauss-Bonnet term does contribute to the
dynamics.  Similarly the cubic order Lovelock term can be added from
six dimensions onwards but it contributes to dynamics only from seven
dimensions onwards. 

We will show that the Schr\" odinger metric is generically not a
solution to the Lovelock equations of motion, however, it exists as a
solution on a co-dimension $1$ locus in the Lovelock coupling space.  We
show that the Schr\" odinger solution exists precisely on the same
locus on which the Lifshitz solution is known to exist\footnote{For closely
  related solutions of Kasner type in the Lovelock theory, see
  \cite{Camanho:2015yqa}.}.  In our computation we restrict ourselves
to the Lovelock terms up to cubic order in the curvature tensor but we
generalize our analysis to arbitrary dimensions.  The co-dimension 1
locus on which we get the Schr\" odinger solution is interesting from
another point of view.  It is known that the Lovelock theories can be
written in terms of the parity preserving Chern-Simons theory.
However, this representation exists only for specific values of the
Lovelock couplings.  The Chern-Simons formulation exists at a point on
this co-dimension $1$ locus on which we find the Schr\" odinger
solutions.  We present these solutions in the Chern-Simons gauge field
forms as well.

The Schr\"odinger solutions are relevant from the point of view of
application to holographically dual condensed matter physics systems.
It then naturally raises a question of relevance of these higher
dimensional solutions to $2+1$ and $3+1$ dimensional condensed matter
systems.  In this regard it is worth pointing out that unlike the AdS
and Lifshitz holography which relates $D$ dimensional theory of
gravity to $D-1$ dimensional field theory, the Schr\"odinger
holography relates $D$ dimensional theory of gravity to $D-2$
dimensional field theory.  Therefore, $4+1$ and $5+1$ dimensional
Lovelock theories are relevant to $2+1$ and $3+1$ dimensional boundary
physics.  Higher dimensional theories can be dimensionally reduced to
lower dimensional theories.  Such higher dimensional theories
typically give rise to scalar-tensor theories of gravity which are
either referred to as Galileon or Horndeski
theories\cite{Nicolis:2008in,Deffayet:2009wt,Deffayet:2009mn,Deffayet:2011gz}.
For example, let us consider $D=d+n+1$ dimensional theory of gravity
with the cosmological constant, the Einstein-Hilbert, the Gauss-Bonnet
term
\begin{equation}
  \label{eq:9}
  S= \int d^Dx\sqrt{-g} \left[ R-2 \Lambda +  a_2 \mathcal{L}_2\right]\ ,
\end{equation}
where, $\mathcal{L}_2$ is the Gauss-Bonnet term. We will dimensionally
reduce it down to $d+1$ dimensions by using an $n$-dimensional compact
manifold $\tilde{K}_n$ such that
\begin{equation}
  \label{eq:10}
  ds^2_D = d\bar{s}^2_{d+1}+ e^{\phi} d\tilde{K}_n^2\ .
\end{equation}
This is a simple but consistent diagonal toroidal compactification
which give rise to one extra scalar degree of freedom, that is the
size of the internal space. All terms with a tilde refers to internal
$n$ dimensional space, while terms with a bar refers to the
$d+1$ dimensional space-time. As we integrate out the internal space the
effective action looks like \cite{Charmousis:2012dw}
\begin{equation}
  \label{eq:11}
  \begin{split}
    \bar{S}_{(d+1)} = \int d^{d+1} x \sqrt{-\bar{g}}\, e^{{n \over 2}\phi}
    &\bigg\{ \bar{R}- 2\Lambda +  a_2 \bar{\mathcal{G}} + {n\over 4}
    (n-1)(\partial \phi \partial \phi) - a_2 n(n-1)
    \bar{G}^{\mu\nu} \partial_{\mu}\phi \partial_{\nu} \phi \\
    & -{a_2 \over 4} n(n-1)(n-2)\left[(\partial \phi \partial \phi) \nabla^2\phi-
    {(n-1)
      \over 4} (\partial \phi \partial \phi)^2\right] \\
    & +e^{-\phi} \tilde{R} \big[ 1+ a_2 \bar{R} +4 a_2
    (n-2)(n-3)(\partial \phi \partial \phi)\big]+ a_2
    \tilde{\mathcal{G}}e^{-2\phi } \bigg\}\ ,
\end{split}
\end{equation}
where, $\bar{\mathcal{G}}$ is $d+1$ dimensional Gauss-Bonnet term,
$\tilde{\mathcal{G}}$ is $n$ dimensional Gauss-Bonnet term and
$\bar{G}^{\mu\nu}$ is the Einstein tensor of $d+1$ dinemsional space.
The effective action written above can be related to the so called
Galileon action, with the Galileon field is realized as the scalar
parametrising the volume of the internal space. As we have reduced
from the Einstein Gauss Bonnet action, in the reduced action all the
terms are up to quartic order in derivatives of the metric or the
scalar or of both, but the equations of motion following from it will
still be of second order. The term of the form
$(\partial \phi \partial \phi) \nabla^2\phi$ is often called the DGP
term\cite{Dvali:2000hr} appearing in the decoupling limit of the DGP
model, and the term of the form $(\partial \phi \partial \phi)^2$ is
the standard Galileon term\cite{Nicolis:2008in}.  As we will argue
that the Schr\"odinger solution exists in a co-dimension 1 subspace of
Lovelock moduli space.  If we restrict to only the Gauss-Bonnet term
then it exists on a subspace which relates $a_2$ to $\Lambda$.  We
will get back to the issue of dimension reduction in this context in
the discussion section.  As long as $n\leq 2$, neither the DGP term
nor the Galileon term appear in the dimensionally reduced theory.  In
addition for Ricci-flat compact spaces the lower dimensional action,
up to the addition of higher derivative curvature terms, has a
familiar form.

This note is organized in the following manner.  We will first give
basics of the Lovelock theory in arbitrary dimensions.  Most of the
information in this section is not new but is useful to fix the
notation.  In the next section we will look at various solutions to
the Lovelock equations of motion.  It is well known that the AdS
solution generically exists for arbitrary values of Lovelock couplings
in any dimension.  This feature is not shared by the Schr\"odinger
solution.  We present the solution for general value of $D$ and in
explicit form for dimensions $D=5,6,7$.  We also comment on the
solutions with anisotropic scaling in spatial direction and their
relation to AdS$\times R$ type solutions. In section 4, we analyse
branches of the AdS solution\cite{Kofinas:2007ns, Jatkar:2015ffa}.
Our interest in presenting this result is to emphasize that the
non-relativistic solutions exists only when the discriminant vanishes.
Degeneracy in the configuration
space\cite{Dotti:2007az,Dehghani:2010kd} has been well studied for
Lifshitz solutions\cite{Kachru:2008yh,Taylor:2008tg}.  We show that
this degeneracy is responsible for unconstrained $z$ for Schr\"
odinger case as well.  Unlike Lifshitz, in the case of Schr\" odinger
solutions this degeneracy extends beyond the Lovelock moduli space.
In this sense our results provide a template for a dynamical exponent $z$
for which all values are equally likely.  Any suitable value of $z$
then can be obtained by either appropriately modifying the couplings
in this theory or by adding new interactions.

It is known that the Lovelock theory in odd space-time dimensions can
be written in the Chern-Simons form and in even space-time dimensions
in the Born-Infeld form exactly when the discriminant vanishes
\cite{Kofinas:2007ns}.  We discuss the relation between vanishing
discriminant and locus of non-relativistic solutions in section 5 and
write down the Schr\" odinger solution in the Chern-Simons gauge field
form in odd space-time dimensions and in the Born-Infeld gauge field
form in even space-time dimensions.  Finally, we point out the
relation with the causality and stability constraints obtained in the
Lovelock theories in higher dimensions \cite{deBoer:2009gx,
  Camanho:2009hu, Camanho:2010ru}.  Finally we summarize our results
and speculate about their applications.  Various technical details are
relegated to Appendix.  Appendix \ref{apdx1} contains details of
Lovelock equations of motion. Appendix \ref{apdx2} recounts details of
AdS and Lifshitz solutions, which are given for the purpose of
comparison with the Schr\" odinger solution. Appendix \ref{apdx3}
contains spin connections and curvature tensors for Schr\" odinger
solution.

\section{The Lovelock Gravity Theory}

Let us consider following action
\begin{equation}
  \label{eq:1}
  I = \frac{1}{16\pi G}\int_{M}d^Dx
  \sum_{p=0}^{[(D-1)/2]} a_p\mathcal{L}_p\ ,
\end{equation}
where $G$ is the $D$ dimensional Newton's constant, $a_p$ are coupling
constants with $a_0=-2\Lambda=(D-1)(D-2)/\ell^2$, $a_1=1$, $a_2$ is
the Gauss-Bonnet coupling etc.\footnote{It is important to note that the mass 
dimensions of various parameters appearing in the action,
eq.(\ref{eq:1}) are as follows 
$[G]=D-2, \ [\Lambda] = [a_0] = 2, \ [a_1]=0, \ [a_2] = - 2, \ [a_3]=
-4, \ \cdots .$
and the parameter $\ell$ has dimensions of length.}, and
$\mathcal{L}_p$ are terms in the Lagrangian density of the Lovelock
action, 
\begin{equation}
  \label{eq:2}
  \mathcal{L}_p = \frac{1}{2^p} \sqrt{-g}
  \delta_{\mu_1\mu_2\cdots\mu_{2p}}^{\nu_1\nu_2\cdots\nu_{2p}} 
R_{\nu_1\nu_2}^{\mu_1\mu_2}\cdots
R_{\nu_{2p-1}\nu_{2p}}^{\mu_{2p-1}\mu_{2p}}\ ,
\end{equation}
where $\delta_{\mu_1\mu_2\cdots\mu_{2p}}^{\nu_1\nu_2\cdots\nu_{2p}}$
is totally antisymmetric product of $2p$ Kronecker deltas normalized
to take values $0$ and $\pm 1$, and hence is completely antisymmetric
in all its upper and lower indices separately. It can also be
considered as the determinant of an $(2p \times 2p)$ matrix whose
$(ij)$-th element is given by $\delta^{\nu_i}_{\mu_j}$.

We are using notation of \cite{Kofinas:2007ns} and the equation of
motion can be written in the compact form as
\begin{equation}
  \label{eq:3}
  E_\mu^\nu = \sum_{p=0}^{[(D-1)/2]}\frac{a_p}{2^p}
  \delta_{\mu\mu_1\mu_2\cdots\mu_{2p}}^{\nu\nu_1\nu_2\cdots\nu_{2p}}
  R_{\nu_1\nu_2}^{\mu_1\mu_2}\cdots 
R_{\nu_{2p-1}\nu_{2p}}^{\mu_{2p-1}\mu_{2p}}=0 \ .
\end{equation}
In this way of writing the Lovelock terms make is obvious that up to
$D=4$ only relevant couplings are $a_0$ and $a_1$.  The term with the
coupling $a_2$ is topological in $D=4$.  In $D=5,6$ the Gauss-Bonnet
term is dynamical.  In these dimensions we will explore the parameter
space spanned by $a_0$ and $a_2$ to find the range of values for which
the Lifshitz solution is possible.  In $D=6$, the term with coupling
$a_3$ can be written down but like the Gauss-Bonnet in $D=4$, this
terms is topological in $D=6$ and hence does not affect the equation of
motion.  However, this term becomes relevant in $D>6$.  We will
explore the three dimensional parameter space spanned by $a_0$, $a_2$
and $a_3$ and find conditions for Lifshitz solutions.

We will start with the study of solution to the $D=5$ equations of motion.
We therefore write the action of pure gravity in $D$-dimensions as
\begin{equation} \label{action1}
 I=\int d^{D} x~ \sqrt{-g}~ \left[R-2 \Lambda + \mathcal{L}_{hd}\right]
\end{equation}
where $\Lambda$ is the cosmological constant and $ \mathcal{L}_{hd}$
is the Lagrangian for the higher derivative terms of the Lovelock form.

As mentioned earlier in $D=5$ space-time dimensions the higher
derivative Lagrangian density contains the quadratic Lovelock term,
also known as the Gauss-Bonnet term, which appears in the Lagrangian
with the coupling constant,  $a_2$,
\begin{equation}\label{deflgb}
\begin{aligned}
 \mathcal{L}_{hd} = &a_2\mathcal{L}_2, ~~ \text{where}, \\
 \mathcal{L}_2= &(R^2-4 R_{\mu\nu}R^{\mu\nu} +
 R_{\mu\nu\rho\sigma}R^{\mu\nu\rho\sigma}) 
\end{aligned}
\end{equation}

For $D=7$ space-time dimensions we will have, in addition to the
Gauss-Bonnet term, the cubic Lovelock term $\mathcal{L}_3$,
\begin{equation} \label{deflll}
 \mathcal{L}_{hd}= a_2\mathcal{L}_2+a_3\mathcal{L}_3
\end{equation}
where $a_3$ is the coupling constant of the cubic Lovelock term.  The
cubic term is explicitly written as
\begin{equation}
\begin{aligned}
 \mathcal{L}_3= &2
 R^{\mu\nu\sigma\kappa}R_{\sigma\kappa\rho\tau}{R^{\rho\tau}}_{\mu\nu}
 + 8 {R^{\mu\nu}}_{\sigma\rho}
 {R^{\sigma\kappa}}_{\nu\tau}{R^{\rho\tau}}_{\mu\kappa}
 +24 R^{\mu\nu\sigma\kappa}R_{\sigma\kappa\nu\rho}R^{\rho}_{\mu} +3
 RR_{\mu\nu\rho\sigma}R^{\mu\nu\rho\sigma} \\
& +24 R^{\mu\nu\sigma\kappa} R_{\sigma\mu}R_{\kappa\nu} + 16
R^{\mu\nu}R_{\nu\sigma}R^{\sigma}_{\mu} -12 RR^{\mu\nu}R_{\mu\nu}+R^3
 \end{aligned}
\end{equation}

The equation of motion that follows from here are as written below,
\begin{equation}\label{eom1}
  G^{(1)}_{\mu\nu}+a_2
  G^{(2)}_{\mu\nu}+a_3 G^{(3)}_{\mu\nu} - \Lambda g_{\mu\nu}=0
\end{equation}
where the explicit forms of $G^{(1)}_{\mu\nu} $, $G^{(2)}_{\mu\nu} $
and $G^{(3)}_{\mu\nu} $ are given in appendix \ref{apdx1}.
The equations of motion in $D=5$ are obtained by setting $a_3=0$ in
(\ref{eom1}).  We will now study specific solutions to the equations of
motion.  

\section{Solutions to the Lovelock Gravity}

In this section we will analyse solutions to the Lovelock gravity
equations of motion.  
In Appendix \ref{apdx2} we will summarise known results about the AdS
and Lifshitz solutions.  In particular the AdS solution is possible
for generic values of the Lovelock couplings $a_2$, $a_3$, etc.  On the 
other hand the Lifshitz solutions exist only on the co-dimension one subspace.
  As we will see below the Schr\" odinger solutions can be obtained only on
the same co-dimension one locus in the parameter space.
Furthermore, as we will see, on this locus the solution
admits arbitrary dynamical scaling exponent $z$, which is also true for 
the Lifshitz solutions. 


\subsection{The Schr\" odinger Solution} \label{schsold5}

We consider the Schr\" odinger space-times as solutions of the higher
derivative Lovelock gravity theories. This is an example of
non-relativistic space-time, known apart from the Lifshitz solution.
We will look for the Schr\" odinger solution to the Lovelock gravity
equations of motion in arbitrary dimensions but we will restrict to
terms cubic in curvature tensor.  Generalization to higher order
Lovelock terms is tedious but straightforward.

The metric ansatz for the Schr\" odinger solution looks like
\begin{equation} \label{Schmet1}
 ds^2= L_{\text{sch}}^2\bigg[-{dt^2 \over r^{2z}}+{dr^2\over
   r^2}+{2\over r^2}dtd\xi+{1\over r^2} \sum_{i=1}^{D-3} dx_i^2\bigg] 
\end{equation}
Note that this metric for the Schr\" odinger solution also has two
parameters, $z$ which is the Schr\" odinger exponent and
$L_{\text{sch}}$ which is the ``Schr\" odinger radius''.  
We will first state the results for arbitrary $D$ but restricting upto cubic
Lovelock terms and then write down explicit expressions for $D=5,6,7$.

The Schr\" odinger space-time solution in general $D$ dimensions
exists subject to following two constraints,
\begin{equation}\label{eq:8}
\begin{split}
\Lambda =& -  {(D-1)(D-2) \over 4L_{\text{Sch}}^2} \bigg( 1 -
(D-3)(D-4)(D-5)(D-6) {a_3\over L_{\text{sch}}^4 }\bigg)\ , \\
a_2 =&{L_{\text{Sch}}^2 \over 2 (D-3)(D-4)}+ {3(D-5)(D-6) \over
  2L_{\text{Sch}}^2}  a_3\ .
\end{split}
\end{equation}
The dynamical exponents $z$ is unconstrained.  If we eliminate
$L_{\text{sch}}$, then it gives one relation between the parameters in
the Lovelock action.  Thus the Schr\" odinger solutions exist on
co-dimension one subspace of the Lovelock moduli space.


In $D=5$ space-time as we have the Gauss-Bonnet term in the Lovelock
action besides the Einstein-Hilbert and the cosmological constant term.
The constraint (\ref{eq:8}) corresponds to 
\begin{equation}
 \begin{split} \label{eq:sch5sol1}
\Lambda = -{3 \over L_{\text{Sch}}^2} ~~  \text{and} ~~ 
 a_2 = \frac{L_{\text{Sch}}^2}{4}\ \Longrightarrow\ a_2\Lambda = -3/4\  .
 \end{split} 
\end{equation}
The non-zero components of the Ricci tensor and the Ricci scalar $R$
for the metric of Schr\" odinger space-time are given by
$R_{tt} = 2 \left(z^2+1\right)/ r^{2 z},\ R_{t\xi} = R_{rr} =R_{x_ix_i}
= -4/r^2$; and  $R= -20/L_{\text{Sch}}^2$.
In $D=6$ space-time the Gauss-Bonnet term is important but the
curvature cubed Lovelock term being a total derivative is not. 
The constraint eq.(\ref{eq:8}) becomes
$\Lambda = -{5/L_{\text{Sch}}^2}$ and $a_2 = L_{\text{Sch}}^2/12$.
We again write down the components of the Ricci tensor
$R_{tt} = \left(2z^2+z+2\right) r^{-2 z}$, $R_{t\xi} = R_{rr} =R_{x_ix_i}
= -5/r^2$
 and the Ricci scalar $R= -(30 / L_{\text{Sch}}^2)$.

In $D=7$ space-time apart from the the Gauss-Bonnet
term, the cubic order Lovelock term will also be important and hence
the action will contain three parameters, $\Lambda$, $a_2$ and $a_3$.
The constraint (\ref{eq:8}) in this case takes the form
\begin{equation} \label{sch7sol1}
 \begin{split} 
   \Lambda = -{15 \over 2L_{\text{sch}}^2}\left(1-{24 \over
    L_{\text{sch}}^4}a_3\right) ~~  \text{and} ~~  
  a_2 = \frac{L_{\text{sch}}^2}{24} + 3  {a_3\over L_{\text{sch}}^2}\ .
 \end{split} 
\end{equation}
The Ricci tensor components are
$ R_{tt} = 2\left(z^2+z+1\right) r^{-2 z},\ R_{t\xi} = R_{rr} =R_{x_ix_i}
= -6/r^2$,
 and the Ricci Scalar becomes $R= -(42 / L_{\text{Sch}}^2)$. 



\subsection{Other Solutions} \label{secaltsol}

Lifshitz solutions\footnote{See eq.(\ref{lifmet2}) for Lifshitz metric in appendix \ref{apdx2}, where we discuss Lifshitz solutions briefly.}, where the time coordinate ($t$) scaled differently
compared to the other coordinates, are known to be solutions of Lovelock gravity.
 As an alternative one can consider
a different version of the Lifshitz solutions where instead of the
time coordinate one of the spatial coordinates may scale differently
compared to others.  We call it ``spatial Lifshitz'' space-time.  More
specifically, we take in $D=5$ dimensions the following metric
\begin{equation}\label{lifaltmet}
 ds^2= L_{\text{Lif}}^2 \left({-dt^2+dr^2 + dx_1^2 
 + dx_2^2 \over r^2}+{dx_3^2 \over r^{2z}}\right) .
\end{equation}
It is obvious from the metric in eq.(\ref{lifaltmet}) that the $x_3$
coordinate scales differently compared to the other coordinates with a
Lifshitz exponent parametrized by $z$.

Following the similar procedure as in the previous two subsections, we
can obtain this ``spatial Lifshitz'' as a solution to the Lovelock action
eq.(\ref{action1}). In $D=5$, for the Gauss-Bonnet case
eq.(\ref{deflgb}), we find the solution to be identical to both the
Schr\" odinger case eq.(\ref{eq:sch5sol1}), and the Lifshitz case
eq.(\ref{Lif5sol2}).  Similarly, for $D=7$ dimensions in the cubic
Lovelock theory we find this ``spatial Lifshitz'' as a solution, again,
identical to both the Schr\" odinger case eq.(\ref{sch7sol1}), and the
Lifshitz case eq.(\ref{lif7sol1}).  In all these cases, this solution
exists at the same point in the coupling space at which the Schr\"
odinger and Lifshitz solution exist.  Thus we see that the special
point continues to be relevant as long as one direction, whether
spatial or temporal, has anisotropic scaling property.  In case of
multiple anisotropic directions also one can show that such solutions
exist at special points in the coupling space but generically this
point is different from the one under consideration.

It is also worth mentioning that the situation here is analogous to
what happens in the case of Schr\" odinger solutions
discussed earlier, namely the dynamical exponent $z$ remains
unconstrained for this solution as well.  This, in particular, allows
us to consider a special case when the dynamical exponent for the
``spatial Lifshitz'' solution vanishes, i.e., $z=0$.  A vanishing $z$
in eq.(\ref{lifaltmet}) corresponds to the metric in the $x_3$
direction being invariant under scaling of the radial coordinate.
Since the metric in the directions transverse to $x_3$ is simple AdS
metric in the Poincar\'e coordinates, we obtain a $AdS_4 \times R$
solution of the form
\begin{equation}\label{ads4rmet}
 ds^2= L^2 \left({-dt^2+dr^2 + dx_1^2  + dx_2^2 \over r^2}+dx_3^2\right) .
\end{equation}
This kind of anisotropic solution was studied earlier in
\cite{Jain:2014vka}, where additional matter in the form of a linearly
varying dilaton was coupled to two derivative Einstein gravity with
negative cosmological constant to construct such anisotropic
solutions.

Finally we would like to mention that Lovelock equations of motion do
not support black brane solutions with either Lifshitz or Schr\" odinger
asymptotic.  

We illustrate this in five dimensions by considering a finite
temperature metric ansatz for the Lifshitz solutions of the following
form
\begin{equation} \label{lifmetfinT}
 ds^2= L_{\text{lif}}^2\bigg[-{f(r)dt^2 \over r^{2z}}+{dr^2\over
   r^2 f(r)}+{1\over r^2} \sum_{i=1}^{3} dx_i^2\bigg] 
\end{equation}
with $f(r) = 1+ c_1 r^{c_2}$, where $c_1$ and $c_2$ are constants.
The equations of motion are solved by this ansatz if $z=1$ and
$c_2 =-2$ with arbitrary $c_1$.  That is
finite temperature black brane solutions with only $AdS$ asymptotic
are allowed.  A similar analysis can be done for Schr\"odinger
solutions and again we find that there are no finite temperature
Schr\"odinger black brane solutions to eq.(\ref{eq:sch5sol1}).   

\section{The Phase-Space of Solutions}

In this section we analyse the parameter space of the higher
derivative theory and understand in some more detail the phase space
of the solutions we obtained in the previous section. We are working
with the action eq.{\eqref{action1}} and the number of 
parameters of the theory depend on $D$.  For $D\leq 4$,
the cosmological constant $\Lambda$ is the only parameter. For $D=5,6$
we have $\Lambda$ and the Gauss-Bonnet coupling constant $a_2$ and
from $D=7$ onwards we also have $a_3$ the cubic term coupling.
While AdS solutions is parametrized only by its radius
$L_{\text{AdS}}$, the Schr\" odinger and Lifshitz solution has two
parameters, radius $L_{\text{Sch}}$, respectively $L_{\text{Lif}}$ and
the dynamical exponent $z$. 
Note that the Schr\" odinger solutions obtained
in the last section and the Lifshitz solutions exist on the locus
which does not pass through the origin of the Lovelock moduli space.
Thus these solutions would cease to exist if we turn off higher
derivative couplings and they cannot be obtained perturbatively in Lovelock
couplings.

 
\subsection{The Phase-Space of Solutions in $D=5$}

The AdS solution in $D=5$ is written in eq.\eqref{ads5sol2}.
One can re-express it as $L_{\text{AdS}}$
being determined in terms of $\Lambda$. 
 \begin{equation}
 \begin{aligned} 
 {1 \over L_{\text{AdS}}^2}={6 \pm \sqrt{36+48 a_2
     \Lambda} \over 24 a_2} 
 \end{aligned} 
 \end{equation}
 which indicates that there are two branches for the AdS solution with
 different AdS-radius. This AdS solution exists when the term within
 the square root is non-negative 
 \begin{equation}
  \Lambda \ge -{3 \over 4 a_2}.
 \end{equation}
 There is one more constraint on the parameters for the existence of
 AdS solution coming from the demand that $L_{\text{AdS}}^2>0$, which
 is $a_2 >0$.

The two branches of the AdS solutions meet at a point in the
2-dimensional phase-space spanned by the parameters $\{\Lambda,~a_2\}$, given by $\Lambda = -3/(4 a_2)$.
The Schr\" odinger solution and the Lifshitz solution in
$D=5$ eq.\eqref{eq:sch5sol1} and eq.\eqref{Lif5sol2}
exist at this point in the phase-space with arbitrary value of
the dynamical exponent $z$.  We will discuss the
relation between unconstrained $z$ and the degeneracy of the
configuration space later in this section.

\subsection{The Phase-Space of Solutions in $D=7$}

The AdS solution in $D=7$ dimensions is given in
eq.\eqref{ads7sol1}. The AdS radius squared $L_{\text{AdS}}^2$ can be
expressed in terms of the parameters $\Lambda,~a_2$ and $a_3$.  From
eq.\eqref{ads7sol1} it is easy to see that one has to solve a cubic
equation for $L_{\text{AdS}}^2$ and the solutions are
\be \label{Ladsrel}
\begin{split}
 L_{\text{AdS}}^2 &= \frac{540 a_2 \Lambda+s\Lambda
   (s\Lambda-15)+225}{3s \Lambda^2} \\ 
 L_{\text{AdS}}^2 &=  -\frac{540 a_2 \Lambda+(s\Lambda+15)^2}{6
   \Lambda^2 s}\pm\frac{i \left(540 a_2 \Lambda- s^2 \Lambda^2
     +225\right)}{2 \sqrt{3} \Lambda^2 s} 
\end{split}
\ee
where 
\be \label{defs}
s^3 = \frac{-135 \left(6 \Lambda \left(15  a_2+6  a_3
      \Lambda-\sqrt{180  a_2  a_3 \Lambda+36  a_3^2 \Lambda^2+50
        a_3-240  a_2^3 \Lambda-75  a_2^2}\right)+25\right)}{\Lambda^3} 
\ee
One can see that the last two roots are complex conjugate of each
other. Now demanding that the term within the square root in
eq.(\ref{defs}) vanishes, {\em i.e.},  
\be
180  a_2  a_3 \Lambda+36  a_3^2 \Lambda^2+50  a_3-240  a_2^3
\Lambda-75  a_2^2 = 0, 
\ee
and also using $\Lambda= {1 \over L_{\text{AdS}}^2}\left(-15 + 180
  {a_2 \over L_{\text{AdS}}^2}  -360 {a_3 \over
    L_{\text{AdS}}^4}\right)$, one obtains a relation between $a_2$
and $a_3$ 
\be
 a_2 = \frac{L_{\text{AdS}}^2}{24} + 3 {a_3\over L_{\text{AdS}}^2}\ 
\Longrightarrow
 \Lambda = -{15\over 2 L_{\text{AdS}}^2} \left(1-{24 a_3 \over
     L_{\text{AdS}}^4}\right)\ .
\ee
These relations are same as those encountered in our study of Schr\"
odinger and Lifshitz solutions in $D=7$.
For these values, the imaginary parts of the second and
third roots in eq.(\ref{Ladsrel}) vanish and they become
equal, whereas the first root remains different,
\begin{equation}
 L_{\text{AdS}}^2 = \frac{-2 \sqrt{60  a_2 \Lambda+25}-5}{\Lambda}, ~
 L_{\text{AdS}}^2 = \frac{\sqrt{60  a_2 \Lambda+25}-5}{\Lambda}, ~
 L_{\text{AdS}}^2 =\frac{\sqrt{60  a_2 \Lambda+25}-5}{\Lambda}.
\end{equation}
Interestingly, there is yet another choice which simplifies solutions
to the cubic equation. Consider the choice $a_2= -5/(12 \Lambda)$,
and set the discriminant equal to zero then the quantity $s$
in (\ref{defs}), vanishes and all three roots
of the cubic become equal to $L_{\text{AdS}}^2=-5/\Lambda$
This point is on the co-dimension one locus and admits
solutions of Schr\" odinger and Lifshitz kind.  

\subsection{Degeneracy of the Configuration Space}
\label{sec:degen-conf-space}

We will now take up the issue of unconstrained
dynamical exponent in the Lifshitz and the Schr\" odinger metrics in
eq.(\ref{lifmet2}) and eq.(\ref{Schmet1})respectively.  It is known in
the literature, see \cite{Dotti:2007az,Dehghani:2010kd}, that in pure
Lovelock theories, unconstrained dynamical exponent $z$ of the
Lifshitz solutions follows from the existence of degeneracy of the
configuration space.  The degeneracy of the configuration space
corresponds to complete arbitrariness in specifying the metric
component $g_{tt}(r)$ on this locus.  Since this metric component is
completely unconstrained, it naturally follows that the
dynamical exponent is not constrained.  While this result is known for
the Lifshitz metrics, we believe our results for the Schr\" odinger
metrics are new. More specifically, in $D=5$ space-time dimensions,
the statement of degeneracy in configuration space amounts to the fact
that a metric ansatz of the following form for the Schr\"odinger
geometry
\be \label{schmetdeg1}
ds^2= L_{\text{sch}}^2\bigg[-f(r){dt^2 \over r^{2z}}+{dr^2\over
   r^2}+{2\over r^2}dtd\xi+{1\over r^2} \sum_{i=1}^{2} dx_i^2\bigg], 
\ee
happens to be a solution for the action in eq.(\ref{action1}) with any
arbitrary choice of the function $f(r)$, at the same locus in the
parameter space, eq.(\ref{eq:sch5sol1}). 

The degeneracy of the configuration space was studied mostly in the
context of the Chern-Simons representation of the Lovelock theory.
Since the Chern-Simons action does not have any free parameters,
this representation exists only at a point in the Lovelock moduli
space.  However, in dimensions $D>6$ the Lifshitz and Schr\" odinger
solutions exist on a subspace which
extends way beyond the Chern-Simons point.  In order to
understand the relation between the degeneracy of the configuration
space and the special locus better, we deform the five dimensional
Lovelock theory by adding $R^2$ and $R_{\mu\nu}R^{\mu\nu}$ terms to
it.  Since neither Lifshitz nor Schr\" odinger solution exist in the
Lovelock moduli space away from this locus, only way to study
dependence of the degeneracy on the couplings is to expand the
coupling constant space by adding new terms.

Let us deform the Lovelock action in $D=5$ (\ref{action1}) by adding
$R^2$ and $R_{\mu\nu}R^{\mu\nu}$ terms. This deformation of the Lovelock
theory is given in terms of two parameters $b_1$ and $b_2$,
\be
\label{actgen1}
I=\int d^{5} x~ \sqrt{-g}~ \left[R-2 \Lambda + a_2 (R^2-4
  R_{\mu\nu}R^{\mu\nu} + R_{\mu\nu\rho\sigma}R^{\mu\nu\rho\sigma})+b_1
  R^2 + b_2 R_{\mu\nu}R^{\mu\nu}\right].
\ee
We are now considering the most general action of gravity up to
quadratic order in curvatures in  $D=5$.  This action now contains $4$
parameters, $\Lambda,~ a_2, ~ b_1$ and $b_2$. 

The Lifshitz solution, eq.(\ref{lifmet2}), as is known in the
literature \cite{Dehghani:2010kd}, occurs at 
\be
\begin{split}
\lambda =& -\frac{3}{L_{\text{Lif}}^2}-b_1\frac{2  z (z+3) (z
  (z+3)+6)}{L_{\text{Lif}}^4} -b_2\frac{  z (z+3)
  \left(z^2+3\right)}{L_{\text{Lif}}^4}, \\
a_2 = &\frac{L_{\text{Lif}}^2}{4}-b_1  (z (z+3)+6)-\frac{b_2}{2}
\left(z^2+3\right).
\end{split}
\ee
When $b_1= b_2=0$ we get back eq.(\ref{Lif5sol2}), the Lifshitz
solution exists for non-zero $b_1,b_2$ but with fixed
dynamical exponent $z$, determined by the parameters of the theory.
This agrees with \cite{Dehghani:2010kd} that the degeneracy in the
configuration space for the Lifshitz solution occurs only at 
\be
\lambda = -\frac{3}{L_{\text{Lif}}^2}, ~~a_2 =
\frac{L_{\text{Lif}}^2}{4}, ~~ b_1=b_2=0. 
\ee
Interestingly, the Schr\"odinger solution with the metric ansatz,
eq.(\ref{Schmet1}) in $D=5$, for a general theory of higher derivative
gravity beyond Gauss-Bonnet theory with action eq.(\ref{actgen1}), is
obtained as 
\be \label{sch5solgen}
\begin{split}
\Lambda =&-\frac{3}{L_{\text{sch}}^2}
-b_1\frac{80}{L_{\text{sch}}^4}+b_2\frac{4  \left(3
    z^2-7\right)}{L_{\text{sch}}^4},\\ 
a_2 =& \frac{L_{\text{sch}}^2}{4}-10 b_1 +b_2  \left(z^2-3\right).
\end{split}
\ee
We recover eq.(\ref{eq:sch5sol1}), as expected, when we put
$b_1=b_2=0$. But, it is interesting to notice that in
eq.(\ref{sch5solgen}), we get a solution with unconstrained $z$ when
$b_2=0$ but $b_1 \neq 0$. Which in turn means, if we go beyond the
Gauss-Bonnet theory and deform it with only $R^2$ term, but with no
$R_{\mu \nu}R^{\mu \nu}$ term, we still obtain a Schr\"odinger
solution with arbitrary dynamical exponent. It is then natural to ask,
if the degeneracy of the configuration space is still present at this
locus in parameter space beyond Gauss-Bonnet point, and it indeed
turns out to be true. More specifically, a Schr\"odinger metric with
the ansatz 
\be
ds^2= L_{\text{sch}}^2\bigg[-f(r){dt^2 \over r^{2z}}+{dr^2\over
   r^2}+{2\over r^2}dtd\xi+{1\over r^2} \sum_{i=1}^{2} dx_i^2\bigg],
\ee  
is a solution to the equations of motion obtained from the action in
eq.(\ref{actgen1}) with $b_2=0$, for 
\be \label{sch5solgen1}
\Lambda =-\frac{3}{L_{\text{sch}}^2} -b_1\frac{80}{L_{\text{sch}}^4},~~ 
a_2 = \frac{L_{\text{sch}}^2}{4}-10 b_1
\ee
with arbitrary $f(r)$.  We thus conclude that the
degeneracy in configuration space and the solutions with arbitrary
dynamical exponent belong to the same locus on the parameter
space.
The special locus in Gauss-Bonnet theory on which both Lifshitz and
Schr\"odinger solutions co-exist also a Chern-Simons description but the
degeneracy of the configuration space of Schr\" odinger solution is
neither confined to Chern-Simons description nor to the Lovelock subspace.
Although, we have carried out the study of degeneracy of the
configuration space in $D=5$ for Gauss-Bonnet theory and
its deformation to more general quadratic curvature 
theories, similar analysis can be done in $D>5$ dimensions. 


\section{Lovelock Gravity as $AdS$ Chern-Simons Gravity 
and Born-Infeld Gravity}

The Lovelock theory has the property that the action has general
covariance and the field equations contain at most two derivatives of
the metric.  We parametrize the Lovelock theory using a set of real
coefficients $a_p, ~p=0,1, \cdots, [D/2]$ which are coupling constants
of the higher derivative terms. It is convenient to adopt the
first order approach, with the dynamical variables being the vielbein,
$e^a = e^a_{\mu} dx^{\mu}$, and the spin connection,
$\omega^{ab} = {\omega^{ab}}_{\mu}~dx^{\mu}$, obeying first order
equations of motion. It is straightforward to solve the vanishing of
the torsion for the connection and eliminate them by writing them in
terms of the vielbeins to obtain the standard second order form in
terms of metric.

The action is constructed as a polynomial of degree $[D/2]$ in
$R^{ab}=(1/ 2) {R^{ab}}_{\mu\nu} ~dx^{\mu}\wedge dx^{\nu}$ 
and
\begin{equation} \label{lifaction}
 I = \frac{1}{16\pi G}\int_{M}d^Dx  \sum_{p=0}^{[D/2]} a_p
 \mathcal{L}_p,\quad \hbox{\rm where},\ 
\mathcal{L}_p = \epsilon_{a_1 \cdots a_D} R^{a_1a_2} \cdots
R^{a_{2p-1}a_{2p}}~e^{a_{2p+1}} \cdots e^{a_D} 
\end{equation}
Imposing the condition that the theory possesses maximum possible
degrees of freedom determines all Lovelock couplings in terms
$\Lambda$ and $G_N$.  The action in odd
dimensions can then be written as a Chern-Simons action with 
$AdS$, $dS$ or Poincar\'e
symmetry\cite{Witten:1988hc,Zanelli:2005sa,Crisostomo:2000bb}, and in
even dimensions as a Born-Infeld like
action\cite{Zanelli:2012px}\footnote{Though having explicit torsion in
  the Lagrangian for $D=4k-1$ is possible with the same requirements,
  we will not consider them here.}.

\subsection{Odd Dimensions: Lovelock Gravity as 
Chern-Simons Gravity} \label{sec5a}

\subsubsection{The Chern-Simons Theory}

It is well known that gravity in $(2+1)$ dimensions can
equivalently be written as a Chern-Simons theory for the gauge groups
$ISO(2,1)$ or $SO(2,2)$, but with no propagating bulk degrees of
freedom. In higher dimensions, $D=2n-1,~n\ge2$, the essential idea for
constructing a Chern-Simons theory is to utilize the fact that there
exists a $2n$-form in $D=2n$,
\be
Q_{2n}({\bf A}) =Tr[{\bf F}^n] = Tr[\underbrace{{\bf F} \wedge {\bf F}
  \wedge \cdots \wedge {\bf F}}_{n -times}]\ .
\ee
This form is closed, i.e. $dQ_{2n} =0$, where ${\bf A}$ is the Lie Algebra
valued connection 1-form 
${\bf A} = A^a_{\mu} {\bf T}_a dx^{\mu}$
and ${\bf F}=d{\bf A}+{\bf A}\wedge{\bf A}$ is the corresponding field strength or curvature 2-form,
with ${\bf T}_a$ being the generators of the Lie algebra $g$ of the
gauge group ${\bf G}$\cite{Troncoso:1999pk}. The fact that $Q_{2n}$ is
closed leads to the existence of a $(2n-1)$-form $L_{CS}^{2n-1}$ such
that \be d L_{CS}^{2n-1} = Q_{2n} = Tr[{\bf F}^n] \ee which can always
be solved as \be L_{CS}^{2n-1} ({\bf A}) = {1 \over (n+1)!} \int_0^1
dt~ Tr\big[{\bf A}(t d{\bf A}+t^2 {\bf A}^2)^{n-1}\big] + \alpha \ee
with $\alpha$ being some arbitrary closed $(2n-1)$-form. This way one
constructs a Chern-Simons Lagrangian $L_{CS}^{2n-1} ({\bf A})$ in
$D=2n-1$ dimensions with an action \be \label{csact} I_{CS} ({\bf A})
=\int_{M_{2n-1}} L_{CS}^{2n-1} ({\bf A}).  \ee

\subsubsection{Connection with the Lovelock Gravity}

In odd dimensions, {\em i.e.}, $D=2n-1$, it was argued that the requirement
of having maximum possible degrees of freedom fixes the Lovelock
coefficients as\cite{Troncoso:1999pk}
\be \label{coeflov}
a_p = {\kappa L^{2p-D} \over {D-2p}} {n-1 \choose p}, \quad 0\le p \le
n-1 
\ee
leaving the action depending on only two parameters, gravitational
constant $\kappa$ and the cosmological constant
$\Lambda$\footnote{Note that $L$ is the length parameter related to
  cosmological constant as $\Lambda = \pm {(D-1)(D-2)\over 2 L^2}$,
  where as the Newton's constant $G_N$ is related to $\kappa$ through
  $\kappa^{-1} = 2(D-2)! \Omega_{D-2} G_N$.}.
The precise connection of Lovelock theories with Chern-Simons gravity
theories in odd dimensions ($D=2n-1$) is that the Lagrangian for the
Lovelock theory can be cast as a Chern-Simons theory for the group
$AdS$. This can be demonstrated through the packaging of the Lovelock
vielbeins $e^a$ and connections $\omega^{ab}$ in the following
connection 1-form as 
\be
W^{AB} = \left[
\begin{array}{ll}
\omega^{ab} & {e^a \over L}\\
-{e^a \over L} & 0
\end{array}\right]
\ee
where the indices $a,b=1,\cdots,D$ and $A,B=1,\cdots,D+1$. Note that
the $A,~B$-indices are raised or lowered with respect to the $AdS$
metric 
\be
\Pi_{AB}=\left[
\begin{array}{ll}
\eta_{ab} & 0\\
0 & -1
\end{array}\right]
\ee
This connection defines a curvature 2-form, also called the $AdS$
curvature, as 
\be \label{defFAB}
F^{AB} = dW^{AB}+{W^A}_{C}\wedge {W_{C}}^{B} = 
\left[
\begin{array}{ll}
R^{ab}+{e^a\wedge e^b \over L^2} & {T^a\over L}\\
-{T^a\over L} & 0
\end{array}\right]
\ee
where $R^{ab}=d\omega^{ab}+{\omega^a}_{c}\wedge {\omega_{c}}^{b}$ is
the curvature 2-form for the 1-form $\omega^{ab}$, 
which is $(2n-1)$-dimensional and not to be confused with the
$2n$-dimensional $AdS$ curvature 2-form $F^{AB}$.  $T^a$ is the
torsion form and setting it to zero corresponds to imposing
torsion-free constraint.  

Next, using the invariant tensor for the $AdS$ group $\epsilon_{A_1
  \cdots A_{2n}}$ along with the $2n$-dimensional AdS curvature
$F^{AB}$ one constructs the Euler form in $2n$-dimension
\be
\mathcal{E}_{2n} = \epsilon_{A_1 \cdots A_{2n}} F^{A_1A_2} \cdots
F^{A_{2n-1}A_{2n}} 
\ee
Using the Bianchi identity for the $AdS$ curvature
$F^{AB}$ we can show that this Euler density is closed,
$d\mathcal{E}_{2n}=0$,
and eq.(\ref{defFAB}) one
can write the $(2n-1)$-form $L_{CS}^{2n-1}$ in terms of the
$(2n-1)$-dimensional curvature 2-form $R^{ab}$ and the vielbeins $e^a$
such that  
\be
L_{CS}^{2n-1} =  \sum_{p=0}^{[D/2]} a_p ~\epsilon_{a_1 \cdots a_D}
R^{a_1a_2} \cdots R^{a_{2p-1}a_{2p}}~e^{a_{2p+1}} \cdots e^{a_D}\ , 
\ee
and $d L_{CS}^{2n-1} = \mathcal{E}_{2n}$. The coefficients $a_p$ are
completely fixed here due to the relation between the
Chern-Simons density and the Euler density and they
turn out to be exactly same as those in eq.(\ref{coeflov}).  The
field equations obtained from the action in eq.(\ref{csact}) are 
\be 
\begin{split}
\epsilon_{a_1a_2a_3\cdots \a_{2n-1}}F^{a_2a_3}\cdots
F^{a_{2n-2}a_{2n-1}}&=0,\\ 
\label{eqCSgrav2}
\epsilon_{a_1a_2a_3\cdots \a_{2n-1}}F^{a_3a_4}\cdots
F^{a_{2n-3}a_{2n-2}} T^{a_{2n-1}}&=0. 
\end{split}
\ee

\subsection{Even Dimensions: Lovelock Gravity as Born-Infeld Gravity}  \label{sec5b}

As we have seen in odd dimensions there are gravity actions which are
invariant not just under Lorentz group but also under some its
extensions, e.g. $AdS$ group $SO(D-1,2)$. On the contrary, this is
not possible in even dimensions, $D=2n$. However the
requirement of having maximum possible number of degrees of freedom
fixes the Lovelock coefficients as\cite{Troncoso:1999pk}
\be
a_p= \kappa {n \choose p}, \quad 0\le p \le n.
\ee
The Lovelock action depends on two constants only, the
gravitational constant and the cosmological constants, and the
Lagrangian, given in eq.(\ref{lifaction}), becomes 
\be
\mathcal{L}= {\kappa \over 2n} \epsilon_{a_1 \cdots a_D} F^{a_1a_2}
\cdots F^{a_{D-1}a_{D}} 
\ee
which is pfaffian of the two form $F^{ab}=R^{ab}+{e^a e^b \over L^2}$
and can be cast in Born-Infeld form\cite{Troncoso:1999pk}
\be
\mathcal{L}= 2^{n-1} (n-1)! \kappa \sqrt{det\left(R^{ab}+{e^a e^b
      \over L^2}\right)}. 
\ee
It is important to note that the two forms $F^{ab}$ are no longer a part
of any $AdS$ curvature.  The field equations in even dimensions take
the form 
\begin{equation}
  \label{eqBIgrav1}
  \begin{split}
    \epsilon_{a b_1 \cdots b_{D-1}}F^{b_1b_2} \cdots F^{b_{D-3}b_{D-2}}
~e^{b_{D-1}} &= 0 \\
\epsilon_{a b a_3 \cdots a_{D}}F^{a_3a_4} \cdots F^{a_{D-3}a_{D-2}}
T^{a_{D-1}}e^{a_D} &= 0.
  \end{split}
\end{equation}

\subsection{Schr\" odinger Space-Time as a Solution to Chern-Simons
  Gravity in $D=5$ Dimensions}

We will now explicitly show that the Schr\" odinger solution obtained
earlier from the Lovelock action can also be seen as a solution to
the Chern-Simons gravity in odd dimensions. We will work in $D=5$
dimensions. The metric in $5$-dimension looks like 
\begin{equation}\label{schmet5}
 ds^2= L_{\text{sch}}^2\bigg[-{dt^2 \over r^{2z}}+{dr^2\over
   r^2}+{2\over r^2}dtd\xi+{1\over r^2} (dx^2+dy^2)\bigg]\ .
\end{equation}
We make the following choice for vielbeins corresponding to the 
metric in eq.(\ref{schmet5}),
\be \label{schviel}
\begin{split}
e^1_t =- {L_{\text{sch}} \over r^z}, ~ e^1_\xi= L_{\text{sch}} r^{z-2},
~ e^3_\xi = L_{\text{sch}} r^{z-2} ,~ e^2_r =e^4_x=e^5_y=
{L_{\text{sch}} \over r}
\end{split}
\ee The spin connections $\omega^{ab}$ and the AdS curvature $F^{AB}$
for the Schr\" odinger metric are listed in appendix \ref{apdx3}.  It
is easy to see that the $AdS$ curvature $F^{AB}$ does indeed satisfy
the field equations 
eq.(\ref{eqCSgrav2}).

\subsection{Relation with Causality and Stability Constraints}
\label{sec:relat-with-unit}

Stability analysis of Lovelock theories in higher dimensions has been
carried out in the past \cite{deBoer:2009gx, Camanho:2009hu,
  Camanho:2010ru}.  These studies derive constraints on the values of
Gauss-Bonnet coupling ($a_2$) and the cubic Lovelock coupling ($a_3$)
by demanding causality and stability condition on the
solutions of the Lovelock theory in $D=7$.  These two conditions are
satisfied in a region in the neighbourhood of the origin of the
($a_2$, $a_3$) plane and at an isolated point, which in our choice of
normalization corresponds to ($a_2=L^2/36,~ a_3= L^4/648$). The
Lovelock parameters ($a_2$, $a_3$) used in this paper are related to the
parameters ($\beta_2$, $\beta_3$) or ($\lambda_1$, $\lambda_2$) used
in \cite{deBoer:2009gx} in the following way 
$a_2 = \beta_2 L^2 = (\lambda_1/12) L^2$ and $a_3=
\beta_3 L^4 = (\lambda_2/ 72) L^4$.
In terms of these parameters the isolated point, mentioned above,
corresponds to ($\lambda_1=1/3$, $\lambda_2=1/9$)\footnote{Note that
  in \cite{deBoer:2009gx} the cosmological constant is taken to be
  $\Lambda=-15/L^2$ and relation between $\beta_i$ and $\lambda_i$ is
  given for $L=1$.  This reduces the coupling parameter space from
  three to two dimensions, therefore the Schr\" odinger or Lifshitz
  solutions exist only a point in the reduced parameter space
  $(\beta_1, \beta_2)$ or $(\lambda_1,\lambda_2)$.}.  It is
interesting to note that this isolated apex point in the phase
diagram, see figure $1$ in \cite{Camanho:2010ru}, is also the same
point where we have the Chern-Simons representation for the Lovelock
theory.  The Schr\" odinger and Lifshitz solutions exist only at this
point in the Lovelock coupling space, which presumably also implies
that they also satisfy the causality and stability constraints.  It
would be interesting to check this explicitly for these solutions.

\section{Discussion}

We studied the coupling constant parameter space of Lovelock gravity
theories in arbitrary dimensions, while restricting our analysis to the
Lovelock terms up to cubic in curvatures.  We demonstrated that
Schr\" odinger solutions exist on co-dimension 1 subspace
in the parameter space.  Similar results for Lifshitz solutions
already exist in the literature.  Interestingly, both the solutions exist on
the same locus.  We found that on this locus, both Schr\" odinger
and Lifshitz exponents were completely unconstrained.  Even if we
couple the Maxwell or Yang-Mills fields to these Lovelock theories,
the Schr\" odinger and the Lifshitz moduli space would continue to be
the same co-dimension 1 subspace.  

As already mentioned earlier in the introduction, Schr\" odinger
holography relates a theory of gravity to field theories living on a
co-dimension 2 subspace. Therefore, $D=5$ and $D=6$ dimensional
Schr\"odinger geometries are directly relevant for studying field
theories with this symmetry in 2+1 and 3+1 dimensions. However, higher
dimensional Schr\"odinger geometries in $D \ge 7$ dimensions are also
relevant for studying field theory systems in lower dimensions, since
the higher dimensional Schr\"odinger space-times can be first
dimensionally reduced to lower dimensions and then we can analyse those
dimensionally reduced theories.  As was pointed out in eq.\eqref{eq:11},
the dimensional reductions do generally lead to effective theories in
the lower dimensions similar to Galileon type theories.  In fact,
in the case when $n=1$ in eq.\eqref{eq:11}, that is starting from $D=d+2$
dimensions we come down to $D=d+1$ dimensions, the effective action
takes even simpler form
 \begin{equation}
   \label{eq:12}
 \begin{split}
 \bar{S}_{(d+1)} = \int d^{d+1} x \sqrt{-\bar{g}}\, e^{\frac{\phi}{2}}
 \bigg\{ \bar{R}- 2\Lambda + a_2 \bar{\mathcal{G}}\ ,
 \bigg\} 
 \end{split}  
 \end{equation}
 and for the case when $n=2$, with simple toroidal reduction it becomes
 \begin{equation}
   \label{eq:13}
 \begin{split}
 \bar{S}_{(d+1)} = \int d^{d+1} x \sqrt{-\bar{g}}\, e^{\frac{\phi}{2}}
 \bigg\{ \bar{R}- 2\Lambda + a_2 \bar{\mathcal{G}} + {1 \over 2}
    \bar{g}^{\mu\nu}\partial_\mu \phi \partial_\nu \phi - 2 a_2
    \bar{G}^{\mu\nu} \partial_{\mu}\phi \partial_{\nu} \phi
 \bigg\}\ .
 \end{split}  
 \end{equation}

 Since these effective actions are obtained are obtained by
 dimensionally reducing the higher dimensional theories, any solution
 of the higher dimensional theories continues to solve the equations
 of motion obtained from the reduced action.
 
 We have seen in section \ref{schsold5} that for a specific value for
 the Gauss-Bonnet coupling constant $a_2$, given in
 eq.\eqref{eq:sch5sol1}, we obtain a Schr\"odinger solution in $D=5$
 dimensions with an unconstrained exponent $z$. Similarly, starting in
 $D=7$ dimensions and for the sake of simplicity allowing only the
 Gauss-Bonnet term, that is assuming $a_3=0$, we can perform a
 toroidal compactification over a $2$ dimensional internal manifold
 and the resulting effective theory in $D=5$ dimension comes with
 action given in eq.\eqref{eq:13}. The Schr\"odinger solution in $D=7$
 dimension, given in eq.\eqref{sch7sol1} with $a_3=0$, also becomes a
 solution to the dimensionally reduced effective theory eq.\eqref{eq:13},
 with unconstrained dynamical exponent $z$ for these particular values
 of the parameters $\Lambda$ and $a_2$. Therefore, from the view point
 of effective lower dimensional theories with the action motivated by
 the dimensional reductions from higher dimensional theories, we can
 take a phenomenological bottom-up approach to study various aspects
 of field theoretical systems with Schr\"odinger symmetries. In this
 regard, our study in this paper provides a template for studying
 non-relativistic field theories with arbitrary dynamical exponent via
 holography, in theories of gravity coupled to matter systems through
 non-trivial but specific values of couplings of the Galileon terms,
 as dictated by eq.\eqref{eq:sch5sol1} and eq.\eqref{sch7sol1}.  For
 $n\leq 2$ there is further simplification because in that case
 neither the DGP term nor the Galileon type terms appear in the $d+1$
 dimensional theory as is evident from eq.\eqref{eq:12} and
 eq.\eqref{eq:13}.  
 Starting from these dimensionally reduced theories, one can then
 deform them appropriately by adding new terms to the action and
 engineer non-relativistic solutions with particular fixed values of
 the dynamical exponent $z$.  Analysis of these deformations leading
 to specific values of $z$ relevant for application to, say, the
 condensed matter systems is beyond the scope of this investigation.

We also pointed that the co-dimension 1 subspace of the Lovelock
theories on which Schr\" odinger and Lifshitz solutions exist also
supports Chern-Simons formulation in odd space-time dimensions and
Born-Infeld formulation in even space-time dimensions.  We cast our
non-relativistic metric in the gauge connection form suitable for
these formulations.  At this point it is interesting to note
that\cite{Zanelli:2005sa,Zanelli:2012px} these gauge connection
formulations have natural super-symmetric extension.  It would be
interesting to explore super symmetric non-relativistic solutions in
the Lovelock theories.

It is known that the Chern-Simons point in the coupling space of the
Lovelock theories is maximally symmetric.  However, most of the
studies at this point are concentrated either on the AdS type
solutions or on the black brane solutions.  Neither of these solutions
can shed direct light on the possible values of the dynamical exponent
$z$ that appears in the Schr\" odinger or Lifshitz solutions. 
Unconstrained dynamical exponent is related to the degeneracy of the
configuration space \cite{Dotti:2007az,Dehghani:2010kd}, which was
studied in the context of Chern-Simons formulation.  We studied
modification of the Gauss-Bonnet theory by doing general deformation
using terms quadratic in curvature and found that the degeneracy of
the configuration space in case of the Schr\" odinger solution is not
confined to the Chern-Simons point but 
extends in the direction orthogonal to the Lovelock moduli space
corresponding to deformation by the Ricci scalar squared term.  Thus
the unconstrained dynamical exponent is a result of the degeneracy and
it has weak dependence on the special locus in the Lovelock moduli
space in the case of the Schr\" odinger solutions.

Another point worth mentioning is that we do not find hyper-scaling
violating solutions anywhere in the Lovelock moduli space, nor do we
find black brane solutions with either Schr\" odinger or Lifshitz type
scaling.  This in turn means it is harder to turn on temperature in
these geometries, however, these shortcomings can be remedied by
deforming away from the Lovelock moduli space.



\paragraph{\large Acknowledgements}: We would like to thank Sujay
Ashok, Anirban Basu, Jyotirmoy Bhattacharya, Sayantani Bhattacharyya,
Anshuman Maharana, K. Narayan, Gautam Mandal, S. Kalyana Rama, Ashoke
Sen and Sandip P. Trivedi for discussion and useful suggestions.  One of
us (DPJ) would like to thanks IMSc for hospitality during the course
of this work. 

\appendix

\section{Appendix}

\subsection{Explicit form of terms in the Equation of
  Motion} \label{apdx1}

In this appendix we write down the explicit forms of
$G^{(1)}_{\mu\nu} $, $G^{(2)}_{\mu\nu} $ and $G^{(3)}_{\mu\nu}$
appearing in the equation of motion eq.(\ref{eom1}). These terms come
from the Einstein term, curvature squared and cubic terms respectively
in the Lovelock action, eq.(\ref{eq:1}).
The term $G^{(1)}_{\mu\nu}= R_{\mu\nu}-{1\over2} g_{\mu\nu} R$ is the
standard Einstein tensor coming from $\mathcal{L}_0$. 
The term $G^{(2)}_{\mu\nu}$, coming from 
$\mathcal{L}_1$ is given by
\begin{equation}
\begin{aligned} 
 G^{(2)}_{\mu\nu} =&~ 2(R_{\mu\sigma\kappa\tau}
 {R_{\nu}}^{\sigma\kappa\tau}- 2
 R_{\mu\rho\nu\sigma}R^{\rho\sigma}-2R_{\mu\rho}R^{\rho}_{\nu}
 +RR_{\mu\nu}) - {1 \over 2} g_{\mu\nu} \mathcal{L}_2 ,
 \end{aligned} 
\end{equation}
and, the third term, $G^{(3)}_{\mu\nu}$, comes from the cubic term
$\mathcal{L}_2$ of the
Lovelock action, 
\begin{equation}
\begin{aligned}
 G^{(3)}_{\mu\nu} =&~ 3 R^2R_{\mu\nu}-12 R_{\mu\rho}R^{\rho}_{\nu}-12
 R_{\mu\nu}R_{\alpha\beta}R^{\alpha\beta}+24
 R_{\mu\alpha}R^{\alpha\beta}R_{\nu\beta} 
 -24 R_{\mu}^{\alpha}R^{\beta\sigma}R_{\alpha\beta\sigma\nu} \\ 
 & +3R_{\mu\nu} R^{\alpha\beta\sigma\kappa} R_{\alpha\beta\sigma\kappa}
 -12 R_{\mu\alpha} R_{\nu\beta\sigma\kappa}
 R^{\alpha\beta\sigma\kappa} -12 R R_{\mu\sigma\nu\kappa} R^{\sigma\kappa}+
 6RR_{\mu\alpha\beta\sigma}{R_{\nu}}^{\alpha\beta\sigma} \\ 
 & +24
 R_{\mu\alpha\nu\beta}R^{\alpha}_{\sigma}R^{\beta\sigma} +24 R_{\mu\alpha\beta\sigma}R^{\beta}_{\nu}R^{\alpha\sigma} +24
 R_{\mu\alpha\nu\beta} R^{\alpha\sigma\beta\kappa} R_{\sigma\kappa} -
 12 R_{\mu\alpha\beta\sigma}R^{\kappa\alpha\beta\sigma}R_{\kappa\nu}
 \\ 
 & -12 R_{\mu\alpha\beta\sigma} R^{\alpha\kappa}
 {R_{\nu\kappa}}^{\beta\sigma} + 24  {R_{\mu}}^{\alpha\beta\sigma}
 R_{\beta}^{\kappa} R_{\sigma\kappa\nu\alpha} -12
 R_{\mu\alpha\nu\beta} {R^{\alpha}}_{\sigma\kappa\rho}
 R^{\beta\sigma\kappa\rho}\\ 
 & -6 {R_{\mu}}^{\alpha\beta\sigma} {R_{\beta\sigma}}^{\kappa\rho}
 R_{\kappa\rho\alpha\nu} -24 {R_{\mu\alpha}}^{\beta\sigma}
 R_{\beta\rho\nu\lambda} {R_{\sigma}}^{\lambda\alpha\rho} - {1 \over
   2} g_{\mu\nu} \mathcal{L}_3\ .
\end{aligned} 
\end{equation}

\subsection{AdS and Lifshitz Solutions in Lovelock Gravity} \label{apdx2}

Here we will list AdS and Lifshitz solutions to the
Lovelock theories in various dimensions. These result are presented
here so that we can compare them with the Schr\" odinger solutions in
the main text. 
The metric ansatz for the AdS solution is 
\begin{equation}\label{adsmet2}
 ds^2= L ^2_{\text{AdS}} \left(-u^2 dt^2+ {du^2 \over u^2} +
   u^2\sum_{i=1}^{D-2} dx_i^2 \right),
\end{equation}
where $L_{\text{AdS}}$ is the AdS radius.  The higher
derivative terms modify the AdS solution by
changing the relation between the AdS radius and the cosmological
constant term in the Lagrangian.  This modification depends on the
dimensions and on the number of Lovelock terms that are turned on.  
The AdS solution in general $D$ dimensions for Lovelock Lagrangians up
to cubic in curvature invariants gives  
\begin{equation}
\begin{aligned}
 \Lambda = -  {(D-1)(D-2) \over 2 L_{\text{AdS}}^2} \bigg[& 1 - { (D-3)(D-4)
 a_2 \over L_{\text{AdS}}^2} -{(D-3)(D-4)(D-5)(D-6) a_3 \over L_{\text{AdS}}^4} \bigg] \ .
 \end{aligned}
\end{equation}

The equations of motion in $D=5$, eq.\eqref{eom1} gives rise to one
condition between the variables $\Lambda$, $L_{\text{AdS}}$ and $a_2$,
\begin{equation}\label{ads5sol2}
 \Lambda = {1 \over L_{\text{AdS}}^2}\left(-6 +12  {a_2 \over
     L_{\text{AdS}}^2} \right)\ \Longrightarrow\ \frac{1}{
   L_{\text{AdS}}^2} = \frac{1}{4a_2}\left(1 \pm 
    \sqrt{1+\frac{4a_2\Lambda}{3}}\right)\ .
\end{equation}
In the Gauss-Bonnet theory there are two
branches of AdS solutions corresponding to two signs of the
square-root in eq.(\ref{ads5sol2}).  These two branches merge when
$a_2\Lambda = -3/4$.  The Lifshitz and Schr\" odinger solutions
exist precisely at these values of the couplings.

In $D=7$ the Lagrangian has three parameters, $\Lambda$,
$a_2$ and $a_3$. The equation of
motion, eq.\eqref{eom1}, is solved by the AdS metric ansatz provided the
following condition is satisfied,
\begin{equation}\label{ads7sol1}
\begin{aligned}
 \Lambda ={1 \over L_{\text{AdS}}^2}\left(-15 + 180 {a_2 \over
     L_{\text{AdS}}^2}  -360 {a_3 \over L_{\text{AdS}}^4}\right)\ .
 \end{aligned}
\end{equation}
This condition can be inverted to write $L_{\text{AdS}}$ in term of
the couplings in the Lagrangian.


We will now consider Lifshitz solutions to the Lovelock equations of
motion.  We consider the metric ansatz for the Lifshitz space-time as
follows\footnote{The general Lifshitz metric has
  $g_{tt} = -L_{\text{Lif}}^2/ r^{{2(D-2)(z-1) \over D-2-\theta}+2}$
  and $g_{rr} = L_{\text{Lif}}^2\ r^{{2 \theta \over D-2-\theta}-2}$
  where, $\theta$ is the hyper-scaling violating
  exponent\cite{Charmousis:2010zz,Iizuka:2011hg,Huijse:2011ef}.  These
  solutions for non-zero $\theta$ do not exist in Lovelock theories.}
\begin{equation}\label{lifmet2}
 ds^2= L_{\text{Lif}}^2 \left(-{dt^2 \over r^{2z}}+  {dr^2 + \sum_{i=1}^{D-2} dx_i^2  \over r^2}\right) .
\end{equation}
The parameter $z$ in eq.\eqref{lifmet2}, which in principle can take any
real value, is called the Lifshitz exponent. 
Asymptotic symmetries are non-relativistic whenever $z\not= 1$.  The
parameter $L_{\text{Lif}}$ is Lifshitz radius.  The Lifshitz solutions
are parametrized by the set of parameters $\{z, ~
L_{\text{Lif}}\}$.


The Lifshitz solution in D dimensional cubic
Lovelock theory exists if \cite{Dehghani:2010kd,Yu:2011zzg}(see also
\cite{Liu:2012yd}) 
\begin{equation}
\begin{aligned}
 \Lambda =& -  {(D-1)(D-2)\over 4L_{\text{Lif}}^2} \bigg( 1 -
 {(D-3)(D-4)(D-5)(D-6) a_3\over L_{\text{Lif}}^4}\bigg)\ , \\
 a_2 =&{L_{\text{Lif}}^2 \over 2 (D-3)(D-4)} + {3(D-5)(D-6) \over
   2L_{\text{Lif}}^2}a_3\ .
 \end{aligned}
\end{equation}


In $D=5$, 
the Lifshitz solution exists whenever following relations are valid,
 \begin{equation}\label{Lif5sol2}
 \begin{aligned} 
 \Lambda = -{3 \over L_{\text{Lif}}^2} ~~  \text{and} ~~ 
 a_2 = \frac{L_{\text{Lif}}^2}{4}
 \end{aligned} 
 \end{equation}
Notice that these relations do not contain $z$, {\em i.e.}, if
(\ref{Lif5sol2}) is satisfied then the Lifshitz solution exists
with arbitrary dynamical exponent $z$. If we eliminate 
$L_{\text{Lif}}$ in eq.(\ref{Lif5sol2}) then the Lifshitz solution
exists only if $a_2\Lambda = -3/4$.


In $D=7$ 
the Lifshitz solution has two conditions which relates three coupling
parameters appearing in the Lagrangian:
 \begin{equation}\label{lif7sol1}
 \begin{aligned} 
 \Lambda = -{15 \over 2 L_{\text{Lif}}^2}\left(1-24 {a_3\over L_{\text{Lif}}^4} \right) ~~  \text{and} ~~ 
 a_2 = \frac{L_{\text{Lif}}^2}{24} + 3 {a_3\over L_{\text{Lif}}^2}\ ,
 \end{aligned}
 \end{equation}
As in five dimensions, we do not get any condition on $z$.  
These values of $\Lambda$ and $a_2$ are also related to the locus in
the parameter space at which three AdS branches merge.  


\subsection{Spin Connections, and $F^{AB}$  for Schr\" odinger Solutions} \label{apdx3}

In this appendix we will follow the discussion of section \ref{sec5a}
and \ref{sec5b} to compute the spin connection $1$-form $\omega^{ab}$
by solving the torsion-free condition
$T^a=de^a+\omega^a_b \wedge e^b=0$.  Using the connection we will then
compute the corresponding curvature $2$-form $R^{ab}$ and finally the
AdS curvature $2$-form $F^{AB}$ defined in eq.(\ref{defFAB}) for the
Schr\" odinger solutions discussed earlier.  

For Schr\" odinger solutions in $D=5$ we consider the 
vielbeins as given in eq.(\ref{schviel}) and compute the 
spin connections explicitly by solving the torsion free condition,
\be
\begin{split}
& \omega^{12}=-\omega^{21}={z \over r^z} dt - r^{z-2} d\xi, \quad 
 \omega^{13}=-\omega^{31}= -{z-1 \over r} dr, \\
 &\omega^{23}=-\omega^{32}=-{z-1 \over r^z} dt+ r^{z-2}d\xi, \quad
 \omega^{24}=-\omega^{42}={1 \over r} dx, \quad
 \omega^{25}=-\omega^{52}={1 \over r} dy 
\end{split}
\ee
The curvature $2$-forms, $R^{ab}$,
can be calculated and the non-vanishing components
are,  
\be
\begin{split}
&R^{12}=L_{\text{sch}}^2 \left[{z^2+(z-1)^2 \over
     r^{z+1}}dt\wedge dr + r^{z-3} dr \wedge d\xi \right],\ R^{13}={L_{\text{sch}}^2\over r^2}dt\wedge d\xi,\\
&R^{23}=-L_{\text{sch}}^2 \left[{2 z(z-1) \over
     r^{z+1}}dt\wedge dr + r^{z-3} dr \wedge d\xi \right],\ R^{24}=-{ L_{\text{sch}}^2\over r^2}dr\wedge dx,\\ 
&R^{14}=L_{\text{sch}}^2\left[  {z \over r^{z+1}}dt\wedge dx
    -r^{z-3} d\xi \wedge dx \right],\ R^{25}=-
  {L_{\text{sch}}^2\over r^2}dr\wedge dy, \\  
&R^{15}= L_{\text{sch}}^2\left[  {z \over r^{z+1}}dt\wedge dy
    -r^{z-3} d\xi \wedge dy \right],\ R^{35}= L_{\text{sch}}^2\left[  {z-1 \over r^{z+1}}dt\wedge
    dy -r^{z-3} d\xi \wedge dy  \right], \\ 
&R^{34}=L_{\text{sch}}^2\left[  {z-1 \over r^{z+1}}dt\wedge dx -r^{z-3} d\xi
    \wedge dx \right],\ R^{45}= -{L_{\text{sch}}^2\over r^2} dx_2
  \wedge dx_3. 
\end{split}
\ee

Finally we compute the $AdS$ curvature $F^{AB}$ defined in
eq.(\ref{defFAB}), note $A,B= 1, \cdots ,6$ and $a,b= 1,
\cdots ,5$ 
\be
\begin{split}
&F^{12} = L_{\text{sch}}^2 {2z(z-1) \over r^{1+z}} dt \wedge dr, \ 
F^{13} = - L_{\text{sch}}^2 {z(z-2) \over r^2} dt \wedge d\xi, \
F^{14} = L_{\text{sch}}^2 {z - 1 \over r^{1+z}} dt \wedge dx, \\
& F^{15} = L_{\text{sch}}^2 {z - 1 \over r^{1+z}} dt \wedge dy, \
F^{34} = L_{\text{sch}}^2 {z - 1 \over r^{1+z}} dt \wedge dx, \ 
F^{35} = L_{\text{sch}}^2 {z - 1 \over r^{1+z}} dt \wedge dy, \\
&F^{23} = -L_{\text{sch}}^2 {2z(z-1) \over r^{1+z}} dt \wedge
dr .
\end{split}
\ee
All the other components, such as, $F^{24},~ F^{25},~ F^{45},~ F^{a6}
= T^a,~ F^{6b} = -T^{b},$ are evaluated to be zero.

\printbibliography

\end{document}